\documentclass{eas}
\usepackage{graphicx}
\begin{document}

\TitreGlobal{Mass Profiles and Shapes of Cosmological Structures}

%%-----------------------------
%%      the top matter
%%-----------------------------
\title{Downsizing Scaling Relations}
\author{Ziegler, B. L.}\address{Institut f\"ur Astrophysik, Friedrich-Hund-Platz 1, 37077 G\"ottingen, Germany}
\author{B\"ohm, A.$^1$}%\address{...}
\author{Fritz, A.$^1$}%\address{...}
\runningtitle{Downsizing Scaling Relations}
\setcounter{page}{23}
% Keep this line, even if the page will be settled afterwards..
\index{Ziegler, B. L.}
\index{B\"ohm, A.}
\index{Fritz, A.}
% Repeat the authors here, this will help to make the final index

%
\begin{abstract}
Faber--Jackson and Tully--Fisher scaling relations for
elliptical and spiral galaxy samples up to $z=1$
provide evidence for a differential behaviour of galaxy evolution with mass.
In compliance with the downsizing scenario, the stellar populations of less
massive galaxies display a stronger evolution than the more massive ones.
For spirals, this may be attributed to a suppressed star formation efficiency
in small dark matter halos.
For ellipticals, star formation must have been negligible at least
during the past $\sim4$\,Gyr in all environments.
\end{abstract}
\maketitle
%
%%-----------------------------
%%      your text
%%-----------------------------
\section{Introduction}

Cold Dark Matter dominated structure formation predicts
that small systems form first which then merge and build up larger
galaxies. If star formation peaks at the epoch of galaxy assembly 
%and declines exponentially thereafter as often assumed for most systems,
then the stellar populations of dwarf galaxies should be
on average older than of giant galaxies. 
% this scenario is at odds with several studies conducted in recent time.
However, the majority of local galaxies in the SDSS show for example a trend 
of increasing age with mass %with a break at around $3\times10^{10}M_o$
(Kauffmann et al. 2003).
%In samples of distant galaxies, similar trends are seen. 
Also, distant spiral galaxies in DEEP exhibit an evolving slope in the 
metallicity--luminosity relation in the sense that smaller galaxies are 
brighter than the larger ones in the past for their metallicities
(Kobulnicky et al. 2003).
%Massive ellipticals in the SXDS at $z\sim1$ are dominated by old stellar 
%populations while less massive ones have extended star formation histories
This phenomenon of a mass-dependent evolution of the stellar population is 
often called the ``anti-hierarchical behaviour'' of the baryons or 
``downsizing'' since it runs opposite to the mass assembly history of galaxies.

\section{Evolution of Spiral Galaxies}

We derive from VLT/FORS spectra 78 and 52 spatially resolved rotation curves 
%(RCs) 
of spiral galaxies with $0.2\le z\le1$ drawn
from the FORS and William Herschel Deep Fields
and combine them with HST/ACS imaging for a Tully--Fisher analysis
(B\"ohm et al. 2004). 
Fitting all distant galaxies with high quality RCs only (63), 
a significantly ($>3\,\sigma$) flatter slope
is found than for the local sample.
Whereas galaxies with
small maximum rotation velocities $v_{\rm max}$ are much brighter than their
local counterparts, the rapid rotators show only very little
evolution. If $v_{\rm max}$ is taken as a measurement of total mass, this
change of slope indicates a mass-dependent evolution of spiral
galaxies.
Fitting optical/NIR colors individually with a chemical enrichment code,
bigger galaxies have on average higher star formation efficiencies and older
mean stellar ages than smaller ones
(Ferreras et al. 2004).

\begin{figure}[t]
   \centering
   \includegraphics[width=6cm]{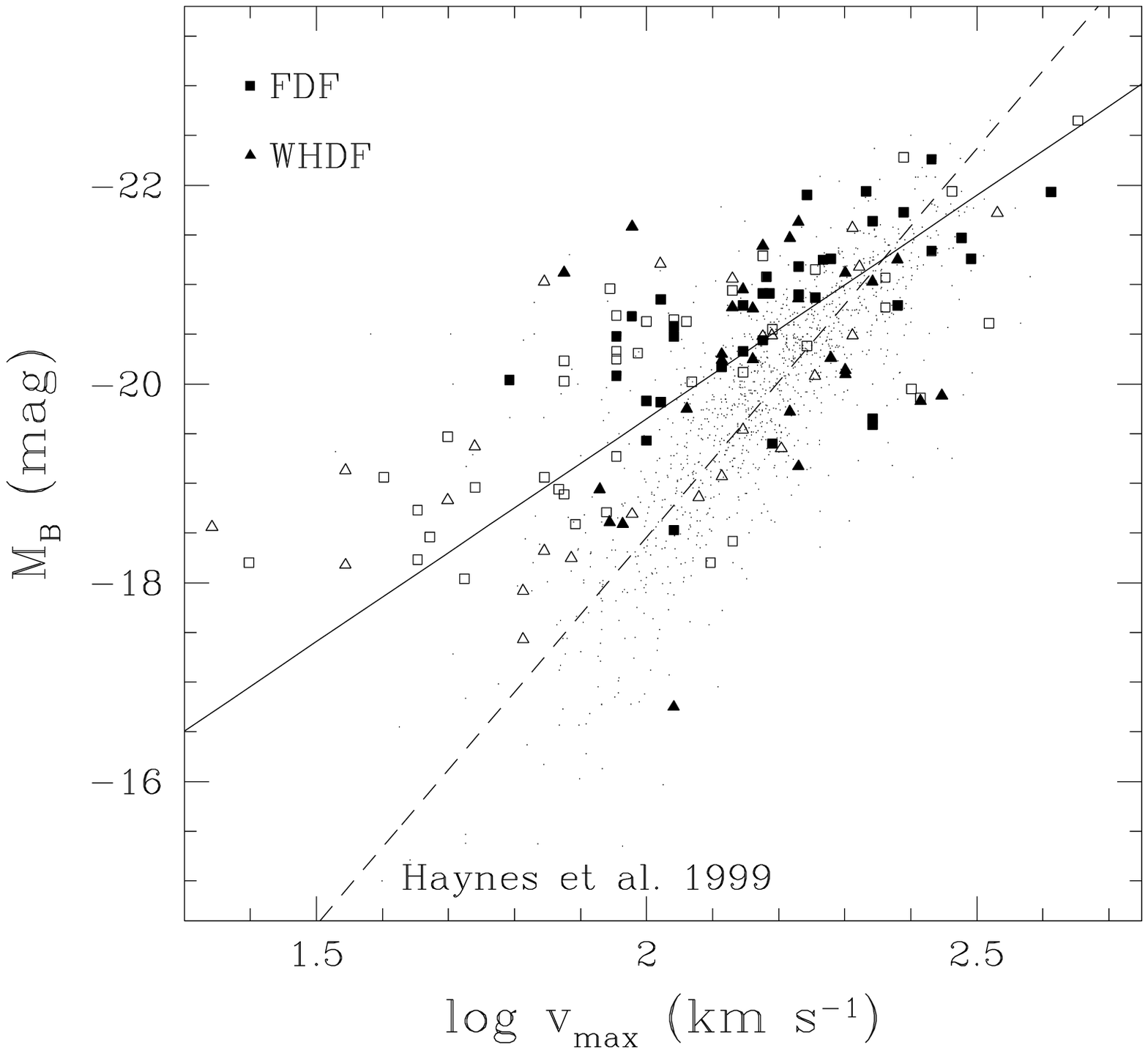}
   \includegraphics[width=5.8cm]{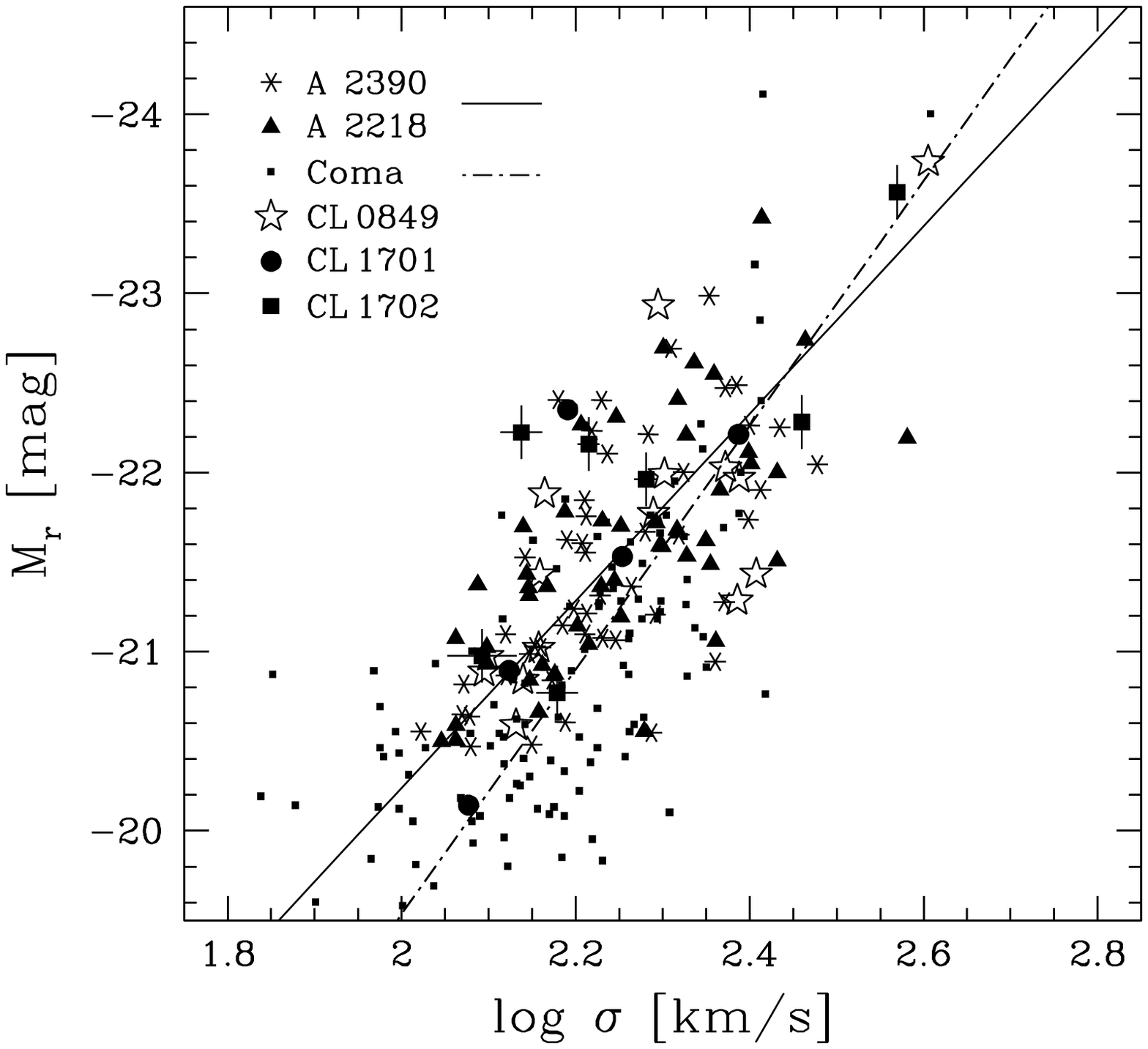}
      \caption{\textsf{Left:} TFR in $B$ of field spirals,
               \textsf{Right:} FJR in $r_{\rm Gunn}$ of cluster ellipticals.}
       \label{zieglerfigure}
   \end{figure}

\section{Evolution of Elliptical Galaxies}

We investigate 96 E and S0 galaxies in the rich clusters Abell\,2218 \& 
Abell\,2390 (Fritz et al. 2005) and 27 early-type galaxies in three poor
clusters that have 1-2 dex lower x-ray luminosities to search for
environmental dependences at $z\approx0.2$.
Both samples display a similar average mild brightening in Gunn $r$ of 
$\sim0.3-0.4^m$ which is compatible with a pure passive evolution of an old 
stellar population. 
Dividing our sample of rich cluster galaxies into halves ($N=48$ each) at 
a velocity dispersion of $\sigma=170$\,km/s, we detect a significant 
difference with less massive ellipticals having a larger offset ($0.6^m$) than
more massive ones ($0.0^m$) from the local Faber--Jackson relation 
of Coma galaxies
(J{\o}rgensen et al. 1995).
%which is in line with the downsizing scenario.

%%-----------------------------
%%      your bibliography
%%-----------------------------
%In preparing the reference list please adhere to the following format.
% Attention should be paid to the order of the items in each reference
% and to the punctuation used. Please see Sect. 4 in the User's Guide
% that comes with the new macro package.

%Bohr, N., Einstein, A., & Fermi, E. 1992, MNRAS, 301, 257 (BEF)
% Curie, M., & Curie, P. 1991, A&A, 248, 612
% de Gaulle, C. 1996, Solar Phys. (Oxford: Oxford Univ. Press)
% Heisenberg, W., & West, C. N. 1993, Australian J. Phys., 537, 36  (Paper III)
% Laurel, S., & Hardy, O. 1994, Active Galactic Nuclei, in The Evolution
% and Distribution of Galaxies, ed. W. Churchill, F. D. Roosevelt, & J.
% Stalin (New York: Wiley), 210

\end{document}